\documentclass[12pt]{iopart}
\usepackage{amsfonts}
\usepackage{amssymb}
\usepackage{txfonts}
\usepackage{graphicx}
\usepackage{bm}
\usepackage{sidecap}

\begin{document}

\title[Non-Markovian Dynamics of Entanglement for Multipartite Systems]{Non-Markovian Dynamics of Entanglement for Multipartite Systems}

\author{Jiang Zhou,Chengjun Wu, Mingyi Zhu and Hong Guo}

\address{CREAM Group, State Key Laboratory of Advanced Optical
Communication Systems and Networks (Peking University) and Institute
of Quantum Electronics, School of Electronics Engineering and
Computer Science, Peking University, Beijing 100871, People's
Republic of China, and Center for Computational Science and
Engineering (CCSE), Peking University, Beijing 100871, People's
Republic of China}
\ead{hongguo@pku.edu.cn}

\begin{abstract}
Entanglement dynamics for a couple of two-level atoms interacting
with independent structured reservoirs is studied using a
non-perturbative approach. It is shown that the revival of atom
entanglement is not necessarily accompanied by the sudden death of
reservoir entanglement, and vice versa. In fact, atom entanglement
can revive before, simultaneously or even after the disentanglement
of reservoirs. Using a novel method based on the population analysis
for the excited atomic state, we present the quantitative criteria
for the revival and death phenomena. For giving a more physically
intuitive insight, the quasimode Hamiltonian method is applied. Our
quantitative analysis is helpful for the practical engineering of
entanglement.
\end{abstract}

 \submitto{\JPB} \maketitle

\section{Introduction}

Entanglement is highly relevant to the fundamental issues of quantum
mechanics and plays a central role in the application of quantum
information \cite{QM1}. In recent years, it has been observed that
two qubits interacting with independent reservoirs experience
disentanglement in a finite time in spite of the asymptotical
decoherence. The corresponding investigations for this phenomenon,
called entanglement sudden death (ESD), give notable results both
theoretically \cite{Yu1,Yu4,Yu2,Z1,Z2,Yu3} and experimentally
\cite{exp1,exp2}. Nevertheless, most of previous works concentrate
on the dynamical evolution of a bipartite system, while the
understanding of how the information is transferred is to be
explored in depth, especially for non-Markovian environment.

Recently, people begin to analyze the dynamics of an extended system
which incorporates two qubits and their respective reservoirs and
successfully explain the process of information transmission in
Markovian regime \cite{multi}. Thanks to these significant
improvements and using the non-perturbative method
\cite{PM1,PM2,PM3}, we extend the study of this composite system
into non-Markovian regime where the reservoir presents some internal
structure.

First of all, we study the entanglement dynamics of two atoms and
two reservoirs. Expectedly, when the atom entanglement is depleted
permanently, it is completely transferred to the reservoir
entanglement irrespective of the non-Markovian memory effect. The
reservoir entanglement can not only exhibit entanglement sudden
birth (ESB) but also ESD. Surprisingly, the revival of atom
entanglement \cite{revival} is not always accompanied by the
disentanglement of reservoirs, and vice versa. Entanglement of atom
pairs can revive before, simultaneously or even after the ESD of
reservoirs and these phenomena are independent of the relative
strength of atom-reservoir interaction.

In addition, we analyze the effective bipartite entanglement and
multipartite entanglement within the system and thus give a
comprehensive interpretation of the information transmission in
non-Markovian regime. Applying the population analysis for the
excited atomic state, we present quantitative criteria, which have
been left obscure for quite a long time, of the occurrence and exact
numbers for revivals of atom entanglement and death of reservoir
entanglement. Our analytic and numerical analysis enables one to
precisely manipulate the entanglement based on the atoms or/and
quantum dots in high Q cavities \cite{exp4,exp5,exp6,exp7} and thus
have potential importance in application.

Finally, by transforming the true mode Hamiltonian \cite{PM2,QO} to
the quasimode form \cite{PM2} which explicitly separates the memory
effect and the damping effect of the reservoir, we give a more
physically intuitive insight into the non-Markovian phenomena.

\section{Theoretical framework}
Under the rotating wave approximation, the truemode Hamiltonian of
the single-body system (atom and its reservoir) is ($\hbar=1$)
\cite{DDH0,LiZhou} \begin{equation}
 H = H_0  + H_{\rm int},
\end{equation}
where
\numparts
\begin{eqnarray}
 H_0  = \omega _0 \sigma _ +  \sigma _ -   + \int_{ -
\infty }^\infty {\textrm{d}\omega _k \omega _k b^\dag  (\omega _k )}
b(\omega _k ),\\  H_{{\mathop{\rm int}} }  = \int_{ - \infty
}^\infty {\textrm{d}\omega _k g(\omega _k )\sigma _ + b(\omega _k )}
+ {\rm{H}}.{\rm{c}}.
\end{eqnarray}
\endnumparts

Here, $ \sigma _ \pm$ and $\omega_0$ are the inversion operators and
transition frequency of the atom, and $b(\omega_k)$,
$b^\dag(\omega_k)$ are the annihilation and creation operators of
the field mode of the reservoir with the eigenfrequency $\omega_k$.

For an initial state of the form $\left| e \right\rangle  \otimes
\left| 0 \right\rangle _r$ with $\left| 0 \right\rangle _r  =
\prod\nolimits_k {\left| {0_k } \right\rangle _r }$, the time
evolution of the single-body system is
\begin{eqnarray}
\label{eq1} \left| {\varphi (t)} \right\rangle  = c_0
(t)\textrm{e}^{ - \rm{i}\omega _0 t} \left| e \right\rangle \left| 0
\right\rangle _r  + \int_{ - \infty }^\infty {\textrm{d}\omega _k
c_{\omega _k } (t)\textrm{e}^{ - \rm{i}\omega _k t} \left| g
\right\rangle \left| {1_k } \right\rangle _r },
\end{eqnarray}
where $\left| {1_k } \right\rangle _r $ is the state of the
reservoir with only one exciton in the $k$th mode. According to the
Schr$\rm \ddot{o}$dinger equation, the equations for the probability
amplitudes take the forms
\begin{equation}
\label{eq2} \rm{i}\dot c_0 (t) = \int_{ - \infty }^\infty {c_{\omega
_k } (t)g(\omega _k )\textrm{e}^{ - \rm{i}(\omega _k - \omega _0 )t}
} \textrm{d}\omega _k,
\end{equation}
\begin{equation}
\label{eq3} \rm{i}\dot c_{\omega _k } (t) = g^* (\omega _k
)\rm{e}^{\rm{i}(\omega _k  - \omega _0 )t} c_0 (t).
\end{equation}
Eliminating the coefficients $c_{\omega _k } (t)$ by integrating
(\ref{eq3}) and substituting the result into (\ref{eq2}), one yields
\begin{eqnarray}\label{eq8}
\dot c_0 (t) =  - \int_0^t {c_0 (t_1 )f(t - t_1 )} \textrm{d}t_1,
\end{eqnarray}
where the correlation function takes the form
\begin{equation}
f(t - t_1 ) = \int_{ - \infty }^\infty  {\rm{d}\omega _k J(\omega _k
)} \rm{e}^{ - \rm{i}(\omega _k  - \omega _0 )(t - t_1 )}.
\end{equation}
Suppose the atom interacting resonantly with the reservoir with
Lorentzian spectral density
\begin{equation}
J(\omega _k ) = \left| {g(\omega _k )} \right|^2  = {W}^2 \lambda
/\pi [(\omega _k  - \omega _0 )^2 + \lambda ^2 ],
\end{equation}
by employing Fourier transform and residue theorem, we get the
explicit form $f(t - t_1 ) = {W}^2 {\rm exp}({ - \lambda \left| {t -
t_1 } \right|})$, where $W$ is the transition strength and the
quantity $1/\lambda$ is the reservoir correlation time.

To construct a concrete physical insight, we adopt a nonperturbative
method called pseudomode approach \cite{PM1,PM2,PM3} in the
following analysis. Equation (\ref{eq8}) can be restated as
\begin{equation}
\dot c_0 (t) =  - \rm{i}Wb(t),
\end{equation}
\begin{equation}
\dot b(t) =  - \lambda b(t) - \rm{i}Wc_0 (t),
\end{equation}
where
\[
b(t) =  - \rm{i}W\int_0^t {c_0 (t_1 )} \rm{e}^{ - \lambda (t - t_1
)} \rm{d}t_1
\]
is the pseudomode amplitude. Typically, there are two regimes
\cite{PM2}: weak-coupling regime ($\lambda>2W$), where the behavior
of the single-body system is Markovian and irreversible decay
occurs, and strong-coupling regime ($\lambda<2W$), where
non-Markovian dynamics occurs accompanied by an oscillatory
reversible decay and a structured rather than a flat reservoir
situation applies. We will limit our considerations to the latter
case.

Define the normalized collective excited state of the reservoir as
\begin{equation}\label{eqdef}
\left| 1 \right\rangle _r  = \int_{ - \infty }^\infty  {c_{\omega _k
} (t)/c_2 (t)\textrm{e}^{ - \rm{i}\omega _k t} } \left| {1_k }
\right\rangle _r \textrm{d}\omega _k.
\end{equation}
one can rewrite (\ref{eq1}) as
\begin{equation}
\label{eq4} \left| {\varphi (t)} \right\rangle  = c_1 (t) \left| e
\right\rangle \left| 0 \right\rangle _r  + c_2 (t) \left| g
\right\rangle \left| 1 \right\rangle _r,
\end{equation}
where
\[
c_1 (t) = \textrm{e}^{ - (\rm{i}\omega _0  + \lambda /2)t}\left[
{\cos \left(\frac{dt }{2}\right) + \frac{\lambda }{d}\sin
\left(\frac{dt }{2}\right)} \right],
\]
with $d=(4W^2-\lambda^2)^{1/2}$. We note that
$c_1(t)=c_0(t)\exp(-\rm{i}\omega_0t)$, $c_2(t)=(1-\left|{c_1 (t)}
\right|^2)^{1/2}$ which can be calculated directly from the
definition (\ref{eqdef}) and
\[
b(t) =  - 2\rm{i}\frac{W}{d}\rm{e}^{ - \lambda t/2} \sin
\left(\frac{dt }{2}\right)
\]
The atom and its reservoir now evolve as an effective two-qubit
system.

Now, we study the joint evolution of two identical single-body
systems initially in the global state
\begin{equation}\label{eqini}
\left| {\phi _0 } \right\rangle  = (\alpha \left| g \right\rangle _1
\left| g \right\rangle _2  + \beta \left| e \right\rangle _1 \left|
e \right\rangle _2 )\left| 0 \right\rangle _{r_1 } \left| 0
\right\rangle _{r_2 },
\end{equation}
where the real non-negative parameters $\alpha$ and $\beta$ satisfy
$\alpha^2+\beta^2=1$ and $i$ ($i=1,2$) denotes the $i$th single-body
system. The evolution of the composite system reads
\begin{eqnarray}
\left| {\phi (t)} \right\rangle  &=& \alpha \left| g \right\rangle
_1 \left| g \right\rangle _2 \left| 0 \right\rangle _{r_1 } \left| 0
\right\rangle _{r_2 }+ \beta \left| {\varphi _1 (t)} \right\rangle
\left| {\varphi _2 (t)} \right\rangle,
\end{eqnarray}
where $\left| {\varphi _i (t)} \right\rangle (i=1,2)$ represents the
single-body evolution and can be determined by (\ref{eq4}).
Employing the density matrix $\rho(t)=\left| {\phi (t)}
\right\rangle \left\langle {\phi (t)} \right|$ and the definition of
concurrence \cite{Concurrence}, we can write down the concurrence of
entanglement for different partitions: $a_1\otimes a_2$, $r_1\otimes
r_2$ and $a_1\otimes r_1$, respectively, \numparts
\begin{eqnarray}\label{eqaa}
C_{a_1a_2} (t) ={\rm max}\{0,2\beta \left| {c_1 } \right|^2 (\alpha
- \beta \left| {c_2 } \right|^2 )\},\\\label{eqrr} C_{r_1r_2} (t)
={\rm max}\{0,2\beta \left| {c_2 } \right|^2 (\alpha
- \beta \left| {c_1 }\right|^2 )\}, \\
\label{eqini1} C_{a_1 r_1 } (t) = 2\beta ^2 \left|
{c_1 } \right|\left| {c_2 } \right|,
\end{eqnarray}
\endnumparts
where $a_i$ ($r_i$) represents the $i$th atom (reservoir).

\section{Dynamic analysis}

We apply the asymptotic analysis to atom and reservoir entanglement.
Suppose, initially, only two atoms are entangled:
$C_{a_1a_2}(0)=2\alpha\beta$ and $C_{r_1r_2}(0)=0$. As
$t\rightarrow\infty$, due to the atom-reservoir interaction for each
subsystem, atom entanglement inevitably vanishes
$C_{a_1a_2}(t\rightarrow\infty)=0$ and reservoir entanglement
necessarily tends to a final value
$C_{r_1r_2}(t\rightarrow\infty)=2\alpha\beta$. This indicates that
the information initially stored in the atoms is completely
transferred to the reservoirs, irrespective of the non-Markovian
memory effect.

\begin{figure}
\centering
 \includegraphics[width=12.cm]{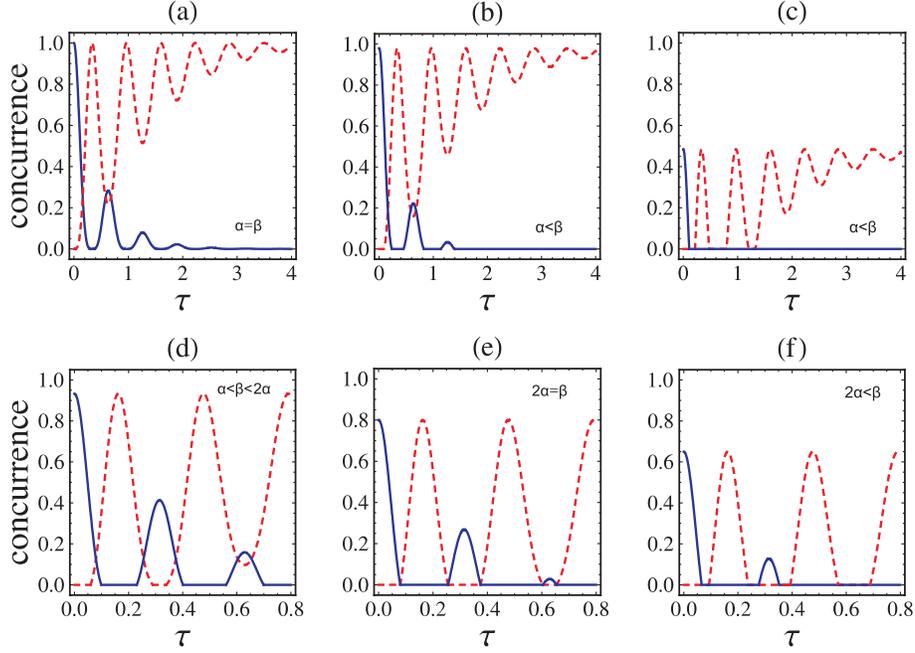}\\
 \caption{Time evolution of concurrence $C_{a_1a_2}$ (solid
 curve) and $C_{r_1r_2}$ (dashed curve) in non-Markovian regime, with
 the initial state being (\ref{eqini}), for the cases of $\lambda/W$ =0.2 (up plots),
 (a) $\alpha=1/\sqrt{2}$, (b) $\alpha=\sqrt{10}/5$,
 (c) $\alpha=1/4$ and $\lambda/W$ =0.1 (bottom plots), (d) $\alpha=2\sqrt{2}/5$, (e) $\alpha=1/\sqrt{5}$, (f) $\alpha=\sqrt{3}/5$.} \label{pic1}
\end{figure}

In Figure \ref{pic1}, we show the time evolution of concurrence
between two atoms (solid curve) and two reservoirs (dashed curve) as
a function of the dimensionless quantity $\tau=\lambda t$ for six
typical values of the parameter $\alpha$, namely,
$\alpha=1/\sqrt{2}, \sqrt{10}/5, 1/4$ ($\lambda/W$ =0.2) and
$2\sqrt{2}/5, 1/\sqrt{5}, \sqrt{3}/5$ ($\lambda/W$ =0.1).

In Figure \ref{pic1} (a), reservoir entanglement performs huge
damped oscillations before it reaches its final value with no ESD or
ESB phenomena and this holds when $\alpha>\beta$. In (b), atom
entanglement can revive from its last death \cite{revival} and
reservoirs can suddenly be entangled in a period of time . In (c),
similar to the Markovian case \cite{multi}, atom entanglement
vanishes permanently after a finite time, while reservoir
entanglement presents sudden birth and death even without the
revival of atom entanglement, which is totally different from the
Markovian case where reservoir entanglement just increases
monotonically up to a stationary value. According to (b) and (c),
the revival of atom entanglement does not necessarily indicate the
sudden death of reservoir entanglement, and vice versa.

Besides these distinctive non-Markovian behaviors, we point out that
entanglement of atom pairs can revive before, simultaneously or even
after the disentanglement of reservoirs, as in Figure \ref{pic1}
(d), (e) and (f). According to the expressions of $C_{a_1a_2}$ and
$C_{r_1r_2}$, (\ref{eqaa}) and (\ref{eqrr}) respectively, the
coincidence of revival and death means $ 2\beta \left| {c_1 }
\right|^2 (\alpha - \beta \left| {c_2 } \right|^2 ) \geq 0 $ and $
2\beta \left| {c_2 } \right|^2 (\alpha - \beta \left| {c_1 }
\right|^2 ) \leq 0 $ where the two equalities hold at the same time
and $2\alpha=\beta$. We can easily prove that revival ahead of
(after) death requires $2\alpha>\beta$ ($2\alpha<\beta$). These
counterintuitive phenomena do not rely on the explicit expressions
of $\left| {c_1 } \right|^2$ or $\left| {c_2 } \right|^2$ and thus
are independent of the relative strength of atom-reservoir
interaction, that is, the ratio $\lambda/W$.

\begin{figure}
\centering
 \includegraphics[width=5 cm]{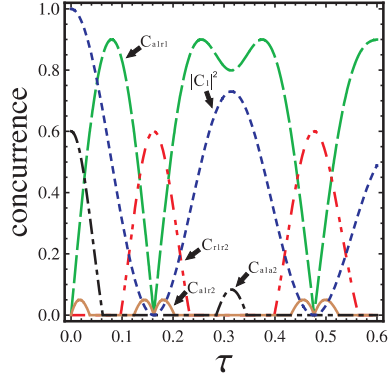}\\
 \caption{Time evolution of concurrence for different partitions in
 strong non-Markovian limit: $C_{a_1a_2}$ (long-short-dashed curve), $C_{r_1r_2}$
 (long-short-short-dashed curve), $C_{a_1r_1}$ (long-dashed curve)
 and $C_{a_1r_2}$ (solid curve), with the initial state
 being (\ref{eqini}) and $\alpha=1/\sqrt{10}$, $\lambda/W
 =0.1$. The short-dashed curve depicts the population $\left| {c_1
 (t)} \right|^2$.} \label{pic2}
\end{figure}

Taking the symmetry of the composite system into account, we analyze
the bipartite entanglement of the partitions: $a_1\otimes a_2$,
$r_1\otimes r_2$, $a_1\otimes r_1$ and $a_1\otimes r_2$ in strong
non-Markovian limit ($\lambda/W=0.1$), as shown in Figure
\ref{pic2}. In the initial period of time, due to the atom-reservoir
couplings, $C_{a_1a_2}$ diminishes and $C_{a_1r_1}$ arises.
According to (\ref{eqini1}), $C_{a_1r_1}$ has a maximal value
$\beta^2$ with $\left| {c_1 (t)} \right|=1/\sqrt{2}$ and if
$2\alpha<\beta$, $a_1\otimes a_2$ and $a_1\otimes r_2$ would already
have been disentangled at that time. With the successive decay of
atoms, the reservoir entanglement emerges and evolves to its maximal
value $2\alpha \beta$ with the two-reservoir state being $\alpha
\left| 0 \right\rangle _{r_1 } \left| 0 \right\rangle _{r_2 }  +
\beta \left| 1 \right\rangle _{r_1 } \left| 1 \right\rangle _{r_2
}$. In Markovian case, the transmission of information comes to an
end.

However, in non-Markovian case, the information transferred to the
reservoirs is fed back to the atoms due to the memory effect, which
means the excited atomic state will arise again. Assuming $\partial
\left| {c_1 (t)} \right|^2 /\partial t = 0$, we get $\sin (dt/2 +
\theta ) = 0$ (valley) or $\tan (dt/2 + \theta ) = d/\lambda$ (peak)
with $\theta =\tan^{-1}( d/\lambda)$. For the peaks, $ \left| {c_1
(t_n)} \right|^2  = {\rm exp}( - 2n\pi \lambda /d) $ ($n \in N$) and
$t_n = 2n\pi /d$. We analyze the behavior of expression $2\beta
\left| {c_1 } \right|^2 (\alpha - \beta \left| {c_2 } \right|^2 )$
between the peak and valley of $\left| c_{1}(t)  \right|^2$ and find
that, the revival of atom entanglement is governed by the inequality
\begin{equation}\label{eqcri}
\alpha  > \frac{{1 - \exp ( - 2n\pi \lambda /d)}}{{\sqrt {1 + [1 -
\exp ( - 2n\pi \lambda /d)]^2 } }}.
\end{equation}
For a given atom-reservoir interaction $\lambda/W$, the occurrence
of revival means
\begin{equation}\label{eqn3}
\alpha  > \frac{{1 - \exp ( - 2\pi \lambda /d)}}{{\sqrt {1 + [1 -
\exp ( - 2\pi \lambda /d)]^2 } }}.
\end{equation}
If $\alpha$ violates the criterion, $C_{a_1a_2}$ fails to exhibit
revival as in Figure \ref{pic1} (c). Furthermore, with a given
initial condition $\alpha$ , the exact number of revivals is
\begin{equation}\label{eqn1}
n_a =[(d/2\pi \lambda )\ln (\beta /(\beta  - \alpha ))],
\end{equation}
where $y=[x]$ is the Gaussian function which represents the maximal
integer smaller than or equal to $x$. The interval of revivals can
be approximately estimated by
\begin{equation}
t_r \approx t_{i+1}-t_i=2\pi/d,
\end{equation}
which is independent of the initial condition $\alpha$ and largely
determined by the transition strength $W$. Therefore, if $\alpha$
satisfies (\ref{eqn3}), the period of revivals is the same. Here, we
note that when $\alpha\geq1/\sqrt{2}$, the atom entanglement
performs damped oscillations and $n$ can be infinite as in Figure
\ref{pic1} (a).

In addition, it is not difficult to derive that the sudden death of
reservoir entanglement is regulated by
\begin{equation}\label{eqcri1}
\alpha  < \frac{{\exp ( - 2n\pi \lambda /d)}}{{\sqrt {1 + \exp ( -
4n\pi \lambda /d)} }}.
\end{equation}
Similar to the atom entanglement, the occurrence of ESD means
\begin{equation}\label{eqn4}
\alpha  < \frac{{\exp ( - 2\pi \lambda /d)}}{{\sqrt {1 + \exp ( -
4\pi \lambda /d)} }}.
\end{equation}
If $\alpha$ is big enough, there is no ESD between the two
reservoirs as in Figure \ref{pic1} (a) and (b). The number of ESD is
\begin{equation}\label{eqn2}
n_r =[ (d/2\pi \lambda )\ln (\beta /\alpha )],
\end{equation}
where $[\cdot]$ is the Gaussian function. Based on (\ref{eqn1}) and
(\ref{eqn2}), we can see that when $\alpha<\beta<2\alpha$, revivals
of atom entanglement are more than deaths of reservoir entanglement
and $2\alpha=\beta$ equal, $2\alpha<\beta$ less. Though this is
visible from the equations, it is not so obvious from our physical
intuition.

Back to our analysis of bipartite entanglement, if the memory effect
is strong enough, that is, the ratio $\lambda/W$ is very small,
$\left| {c_1 (t_{\rm 1})} \right|={\rm exp}(-\pi \lambda /d)$ is
bigger than $1/\sqrt{2}$ and $C_{a_1r_1}$ will rise again to its
maximal value. If $\alpha$ satisfies (\ref{eqn3}) and (\ref{eqn4}),
the reservoir entanglement will be disentangled and the atom
reservoir entanglement will revive, as in Figure \ref{pic2}. This
means the information is transferred from two reservoirs to each
subsystem and then to two atoms. This depletion-feedback process
continues for some time with a damping of amplitudes because the
memory effect is finite and the atoms will decay inevitably.

\begin{figure}
\centering
 \includegraphics[width=12 cm]{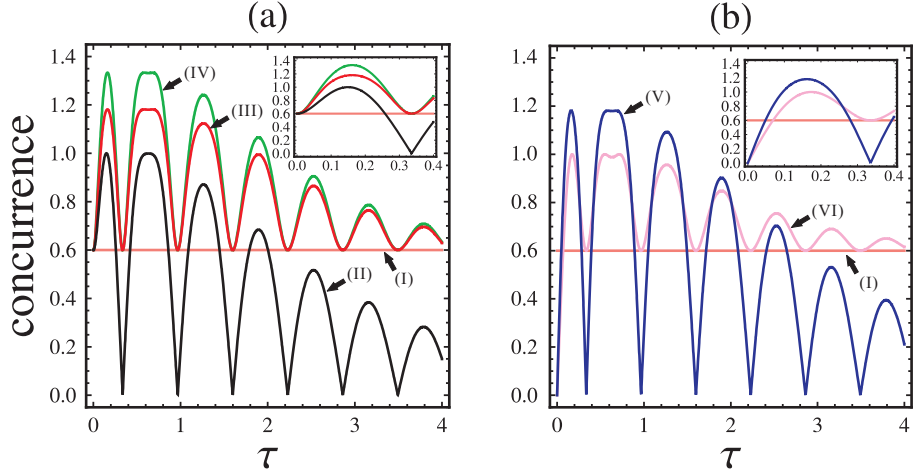}\\
 \caption{Time evolution of entanglement for different partitions in non-Markovian regime: (I) $(a_1\otimes r_1)\otimes (a_2\otimes r_2)$,
 (II) $a_1\otimes (r_1\otimes a_2\otimes r_2)$, (III) $(a_1\otimes r_2)\otimes (a_2\otimes r_1)$, (IV) $a_1\otimes r_1\otimes a_2\otimes r_2$,
 (V) $(a_1\otimes a_2)\otimes (r_1\otimes r_2)$, (VI) $r_1\otimes (a_1\otimes a_2\otimes r_2)$, with the initial state
 being (\ref{eqini}) and $\alpha=1/\sqrt{10}$, $\lambda/W
 =0.2$.}
 \label{pic3}
\end{figure}

Entanglement for other partitions is shown in Figure \ref{pic3}.
Curve (IV) depicts the multipartite entanglement between the four
effective qubits, which is defined by multipartite concurrence $C_N$
\cite{multipartite}. Other bipartite entanglement is obtained
through \textit{I}-concurrence \cite{qudit}, as curves (I)-(III),
(V) and (VI), which coincides with the concurrence in the pure
two-qubit case. We compile the partitions initially entangled in
Figure \ref{pic3} (a) and disentangled in (b). Here, we note that
the partition $(a_1\otimes r_1)\otimes (a_2\otimes r_2)$ has
constant entanglement, specifically $2\alpha\beta$, despite of the
memory effect and information transmission, which serves as a
benchmark of our entanglement analysis. $C_N$ has the same value at
$t=0$ and $t\rightarrow \infty$, showing complete entanglement
transfer from atom pairs to reservoirs, as the asymptotic analysis
indicates. Although long time evolution of entanglement has the same
tendency with the Markovian case \cite{multi}, transient
entanglement for different partitions endures huge damped
oscillations due to the feedback of information.

Our simulation conditions $\lambda/W=0.2$ and $0.1$ can be well
realized within the current experimental level \cite{exp3}. A more
intuitive interpretation of these results will be given in detail in
the following section.

\section{Explanations via quasimode Hamiltonian}
\begin{figure}
\centering
 \includegraphics[width=5 cm]{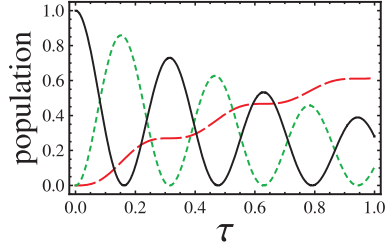}\\
 \caption{Time evolution of population in strong non-Markovian regime
 for quasimode Hamiltonian: $\left| {c_a (t)} \right|^2$ (solid
 curve), $\left| {c_m (t)} \right|^2$ (short-dashed curve) and
 $\left| {c_r (t)} \right|^2$ (long-dashed curve), with the initial
 state being $\left| e \right\rangle \left| 0 \right\rangle _m \left|
 0 \right\rangle _{r'}$ and $\lambda/W =0.1$.} \label{pic4}
\end{figure}

Memory effect plays an important role in our analysis of
non-Markovian dynamics. To separate it from the damping effect, we
convert the true mode Hamiltonian with Lorentzian spectral density
into the quasimode form. Applying the method in \cite{PM2}, the
quasimode Hamiltonian of the single-body system can be given by
\begin{eqnarray}\label{eq6}
H = H_0  + H_{{\rm memory}}  + H_{{\rm damping}},
\end{eqnarray}
with
\numparts
\begin{eqnarray}
H_0  = \omega _0 \sigma _ +  \sigma _ -   + \omega _0 a^\dag  a +
\int_{ - \infty }^{  \infty } {\Delta c^\dag  (\Delta )c(\Delta )}
\textrm{d}\Delta,\\
H_{{\rm memory}}  = W(\sigma _ +  a + \sigma _ - a^\dag
),\\
\label{eq7}
H_{{\rm damping}}  = (\lambda /\pi )^{1/2} \int_{ -
\infty }^{ \infty } {(a^\dag  c(\Delta ) + a} c^\dag (\Delta
))\textrm{d}\Delta,
\end{eqnarray}
\endnumparts
where $c^\dag  (\Delta ),c(\Delta)$ are the creation and
annihilation  operators of the continuum quasimode with frequency
$\Delta$ and other parameters are the same as before. The particular
conversion relations are given in the Appendix in \cite{LiZhou}.
Analogous to the procedure in section II, the evolution of the
single-body system is
\begin{eqnarray}
\left| {\psi (t)} \right\rangle  &=& c_a (t)\left| e \right\rangle
\left| 0 \right\rangle _m \left| 0 \right\rangle _{r'}  + c_m
(t)\left| g \right\rangle \left| 1 \right\rangle _m \left| 0
\right\rangle _{r'}
\\\nonumber& &+ c_r (t)\left| g \right\rangle \left| 0 \right\rangle _m \left| 1
\right\rangle _{r'},
\end{eqnarray}
with $c_a(t)=c_1(t)$, $c_m(t)=b(t){\rm exp}(-\rm{i}\omega_0t)$ and
$c_r (t) = (1 - \left| {c_a (t)} \right|^2  - \left| {c_m (t)}
\right|^2 )^{1/2}$. This means the pseudomode in section II is just
the discrete quasimode and the quasimodes (both discrete and
continuum) are our previous reservoir. We plot the time evolution of
the population in Figure \ref{pic4} and explain how the memory
effect and damping effect induce $\left| {c_1 (t)} \right|^2$ to
perform damped oscillations.

According to (\ref{eq6})-(\ref{eq7}), the atom only interacts with
the discrete mode and their coupling coefficient is just the
transition strength $W$; whereas, the discrete mode interacts with a
set of continuum modes and their coupling strength contains only the
constant width of Lorentzian spectral density $\lambda$. If we let
the atom and the discrete mode be a new system and the continuum
modes be the external environment, the behavior of the new system
will be exactly Markovian. Therefore, transmission of the exciton
from atom to reservoir is a two-step process: first, a photon is
created in a discrete (cavity) mode via the atom-discrete mode
interaction; second, this photon is annihilated and a photon is
created in a continuum (external) mode via the discrete-continuum
mode coupling. The memory effect stems from the finite life span of
the photo in the cavity \cite{PM2,PM3}. Thus, the reabsorbing
phenomenon that causes the oscillatory entanglement, as shown in
Figure \ref{pic1} and \ref{pic2}, only exists in the first step. In
Figure \ref{pic4}, the alternative peaks and valleys of $\left| {c_a
(t)} \right|^2$ and $\left| {c_m (t)} \right|^2$ demonstrate that
the energy of exciton exchanges between the two states $\left| e
\right\rangle \left| 0 \right\rangle _m \left| 0 \right\rangle
_{r'}$ and $\left| g \right\rangle \left| 1 \right\rangle _m \left|
0 \right\rangle _r $ periodically.

Assuming $\partial \left| {c_m (t)} \right|^2 /\partial t = 0$, we
get $\sin (dt/2) = 0$ (the valley) or $\tan (dt/2) = d/\lambda$ (the
peak). The valley of $\left| {c_m (t)} \right|^2$ coincides with the
peak of $\left| {c_a (t)} \right|^2$ at $t = 2n\pi /d$ when
$\partial \left| {c_r (t)} \right|^2 /\partial t =0$, which makes
$\left| {c_r (t)} \right|^2$ act like a staircase curve. This means
that if the exciton only exists in the atom, the damping process
does not happen. The damping effect happens only when the photon
escapes into the continuum modes and never comes back. In our case,
there is no mechanism, such as the dipole-dipole interaction
\cite{LiZhou}, to protect the exciton from escape. So the damping
effect is unavoidable and the memory effect is just a finite-time
phenomenon, which indicates that atom entanglement can never be
completely reconstructed and the information will be totally
transferred to the reservoirs, which fits the results in Figure
\ref{pic1} and \ref{pic3}.

\section{Conclusions}
We study the dynamical evolution of a couple of two-level atoms
interacting with independent structured reservoirs. We find that
reservoir entanglement exhibits sudden birth, death and revival
phenomena. The revival of atom entanglement does not necessarily
indicate the disentanglement of reservoirs, and vice versa. Applying
the quantitative analysis, we derive the criteria for the revival
(death) phenomena and prove that the atom entanglement can revive
before, simultaneously or after the sudden death of reservoir
entanglement, which is independent of the relative strength of
atom-reservoir interaction. Besides, by studying the bipartite and
multipartite entanglement for different partitions, we present a
comprehensive interpretation of the information transmission within
this composite system in non-Markovian regime. Our results and
conclusions are desirable in the implementation of various optical
schemes \cite{exp4,exp5,exp6,exp7} for the preparation and
manipulation of entanglement.

\section*{Acknowledgement}
This work is supported by the Key Project of the National Natural
Science Foundation of China (Grant No. 60837004), and the Open Fund
of Key Laboratory of Optical Communication and Lightwave
Technologies (Beijing University of Posts and Telecommunications),
Ministry of Education, People's Republic of China.

\end{document}